\documentclass[%
 reprint,
 amsmath,amssymb,
 aps,
]{revtex4-2}

\usepackage{graphicx}
\usepackage{dcolumn}
\usepackage{bm}

\begin{document}

\preprint{APS/123-QED}

\title{Bondi news in de Sitter
		space-time}

\author{Djeyrane-Sophie Erfani}
 \altaffiliation{j\_erfaniharami@sbu.ac.ir}
 \address{Department of Physics, Shahid Beheshti University, G.C., Evin, Tehran 19839,Iran} 



\date{\today}

\begin{abstract}
	Observations have shown that a model with a positive cosmological constant is more appropriate to describe our universe. The aim of this manuscript is to study the gravitational fields in de Sitter space-time. To achieve this goal, de Sitter metric is written in Bondi Sachs form and the Einstein equations are solved for this form of the metric in presence of $\Lambda$. On the other hand, a null tetrad is defined for the de Sitter metric and its form in Bondi gauge is obtained by using the general transformation. Then a new result for the Bondi news in de Sitter background is obtained, using the coordinate independent notion of Bondi news which was first presented by Penrose in  1963.
\end{abstract}

\maketitle


\section{\label{sec:level1}introduction}

In 2016 the first ever direct detection of gravitational waves has been celebrated  \cite{abbott2016gw150914, abbott2016observation}. Gravitational waves were one of the predictions of Einstein's field equations. 	Actually in 1917, Einstein had isolated the
radiative modes of the gravitational field in the linear approximation and derived the known quadruple formula. For many years, discussions had been made to define whether they were mathematical artifacts or  real physical phenomenons. These doubts and debates extended till the 1960 when Bondi represented a new way to study gravitational waves \cite{bondi1960gravitational}. Afterwards in 1962, Bondi, Van der Burg and Metzner established in greater detail a framework to study
expansions of axi-symmetric isolated systems so the gravitational
radiation in full Einstein theory would be comprehensible \cite{bondi1962gravitational}. As gravitational waves move with the speed of light, they studied the expansion of the metric when one goes to the null infinity. Gravitational waves can be described in the TT gauge, the gravitational news, $N_{\mu\nu}$, or the Weyl spinor, $\psi_4$. In this article our focus is on gravitational news. Actually,  gravitational radiation caused trace-free,
transverse tensor field at null infinity, $N_{\mu\nu}$ that is known as Bondi news. This traceless tensor can describe the gravitational field \cite{bondi1962gravitational,sachs1962asymptotic}. The general form of the news $N_{A B}=1/2 \partial_{u} C_{A B}(u,x^A)$ is dependent on Bondi coordinates $u$ and $x^A$ and it is just valid  in this coordinate system. Actually a more useful notion of Bondi news  based on a
null tetrad has been defined \cite{penrose1963asymptotic,bishop1997high}, which is used in this article. 

However, as observations showed, our universe is accelerating \cite{riess1998observational, aldering1998measurements,perlmutter1999measurements} so a model with a positive cosmological constant may be a better candidate to describe it. As de Sitter space-time is the solution of Einstein's equations in the presence of a positive cosmological constant, it would be useful to study the expansion of the gravitational field in de Sitter background. Over the years many studies have been done about gravitational waves in de Sitter background \cite{ashtekar2014asymptotics,ashtekar2015asymptotics,he2015new,szabados2015positive, bishop2016gravitational,ashtekar2016gravitational,chrusciel2016cosmological,saw2016mass,he2018relationship, saw2018bondi}.
It seems that the simplest way to study the de Sitter asymptotic symmetries is to follow the same route as in asymptotically flat case.
But unfortunately, $\Lambda$ no matter, how small it is, complicates everything. As an example, the null infinity of an asymptotically de Sitter space-time is no longer null. So it is not as simple as it seems at the first glance. 
In fact,  in presence of a positive constant it is not possible to do the same thing as by using these boundary
conditions gravitational waves do not carry away de Sitter charges across future
null infinity \cite{ashtekar2014asymptotics,ashtekar2015asymptotics}. Therefore one can conclude that these boundary conditions are too restrictive. However, it is shown that mixed Dirichlet-Neumann boundary conditions can be used \cite{compere2019lambda}. So it would be useful to write the de Sitter metric in the Bondi gauge and obtain the Bondi news.

In this manuscript a null tetrad is defined for the de Sitter space-time in $(u,\rho,x^A)$ coordinates  based on the work of Saw \cite{saw2016mass} who has obtained a null tetrad for the de Sitter space-time in "coordinate-like Eddington-Finkelstein-type" coordinates \cite{saw2017mass}. Then by using the general transformation we obtain the tetrad in Bondi gauge. Afterwards we will get the Bondi news by using a notion that is based on a null tetrad \cite{penrose1963asymptotic, bishop1997high}.

Bondi-sachs metric, Bondi gauge and infinitesimal transformations that preserve this form of the metric are reviewed in section \ref{sec:level2}. The conformal form of the metric is considered in section \ref{sec:level3}, Einstein equations are solved in section \ref{sec:level4} after the work of Compère et.al and Poole and his collaborators \cite{compere2019lambda,poole2019ds4}. A null tetrad for the de Sitter metric is defined in section \ref{sec:level5} and its form under general transformation is obtained in section \ref{sec:level6}.  Finally the Bondi news is achived in section \ref{sec:level7}.

\section{\label{sec:level2}The Bondi-Sachs metric}

	If one foliates the space-time to $u=constant$ hypersurfaces,
	the Bondi-Sachs coordinates $(u,r,x^A)$ can be used where $u=constant$ hypersurfaces  are null.  This implies that $g_{11}=0$ as the normal vector, $\partial^{\mu}u$, should satisfy $\partial^{\mu}u \partial_{\mu}u=0$. Then we must have $\Gamma^{0}_{11}=\Gamma^{2}_{11}=0$ that results $r^4\sin^2\theta=g_{22}g_{33}$. So the line element takes the form \cite{bondi1962gravitational,sachs1962gravitational,sachs1962asymptotic}
	\begin{equation}
	\begin{aligned}
		\label{1}
		ds^2=&-e^{2\beta}\frac{V}{r}du^2-2e^{2\beta}du dr\\
		&+r^2h_{AB}(dx^A-U^Adu)(dx^B-U^Bdu)
	\end{aligned}
		\end{equation}
	where $x^{A}=(\theta, \phi)$ and $\beta,\: V,\: h_{AB}$ are functions of $(u,x^{A})$. Also the contravariant form can be written as \cite{madler2016bondi}
		\begin{eqnarray}
		g^{\mu\nu}=
		\begin{bmatrix}
			0&-e^{-2\beta}&0\\
			-e^{-2\beta}&\frac{V}{r}e^{-2\beta}&-e^{-2\beta}U^A\\
			0&-e^{-2\beta}U^A&r^{-2}h_{AB}
		\end{bmatrix}.
	\end{eqnarray}

	$g_{AB}=r^2h_{AB}$ has to satisfy the gauge condition
	\begin{eqnarray}
		\label{eq:6}
		\partial_{r}\left(\frac{\operatorname{det}\left(g_{A B}\right)}{r^{4}}\right)=0 \Leftrightarrow \operatorname{det}\left(g_{A B}\right)=r^{4} \chi\left(u, x^{A}\right).
	\end{eqnarray}
	 $\chi(u, x^{A})$ is an arbitrary parameter. Infinitesimal diffeomorphisms that preserve the Bondi gauge are generated by $\xi^{\mu}$ and have the following relation
\begin{eqnarray}
	 	&\mathcal{L}_{\xi} g_{r r}=0, \quad \mathcal{L}_{\xi} g_{r A}=0, 
	 	\\
	 	&\quad g^{A B} \mathcal{L}_{\xi} g_{A B}=4 \omega\left(u, x^{A}\right)= \frac{4}{r} \xi^{r}+\chi^{-1} \mathcal{L}_{\xi} \chi. \nonumber
\end{eqnarray}
	 This relation is obtained by using \ref{eq:6}. Then one can write the following relation \cite{compere2019lambda}
	 \begin{eqnarray}
	 		&\xi^{u} &=f, \\
	 		&\xi^{A} &=Y^{A}+I^{A}, \quad I^{A}=-\partial_{B} f \int_{r}^{\infty} d r^{\prime}\left(e^{2 \beta} g^{A B}\right),\nonumber \\
	 		&\xi^{r} &=-\frac{r}{2}\left(\nabla_{A} Y^{A}-2 \omega+\nabla_{A} I^{A}-\partial_{B} f U^{B}+\frac{1}{2} f g^{-1} \partial_{u} g\right),\nonumber
	 \end{eqnarray}
	 where $\nabla_{A}$ is the covariant derivative which is related to $g_{AB}$.

\section{\label{sec:level3}The conformal metric}

Rewriting the line element \ref{1} with $\rho=1/r$, one has

\begin{equation}
\begin{aligned}
	\label{eq:3}
	d s^{2}=&\rho^{-2}(-(e^{2 \beta}(\rho^{3}V-h_{A B} U^{A} U^{B}) d u^{2}+2 e^{2 \beta} d u d \rho \\
	&-2 h_{A B} U^{B} d u d x^{A}+h_{A B} d x^{A} d x^{B}).
	\end{aligned}
\end{equation}
It is known that by multiplying a conformal factor by the physical metric ($\hat{g}_{\mu\nu}=\Omega^2 g_{\mu\nu}$) one can manage to attribute infinity to the boundary of a larger space-time, $(\hat{M},\hat{ g}_{\mu\nu})$ \cite{wald2010general}. Thus taking the conformal factor ($\Omega=\rho$), \ref{eq:3} becomes
\begin{equation}
\begin{aligned}
	d \hat{s}^{2}=\rho^{2}d s^{2}=&-e^{2 \beta}(\rho^{3}V-h_{A B} U^{A} U^{B}) d u^{2}+2 e^{2 \beta} d u d \rho\\
	&-2 h_{A B} U^{B} d u d x^{A}+h_{A B} d x^{A} d x^{B}.
	\end{aligned}
\end{equation}

The metric connection is not invariant under conformal transformations

\begin{equation}
	\hat{C}_{\alpha \beta}^{\gamma}=C_{\alpha \beta}^{\gamma}+\frac{\delta_{\alpha}^{\gamma} \partial_{\beta} \Omega+\delta_{\beta}^{\gamma} \partial_{\alpha} \Omega-\hat{g}_{\alpha \beta} \hat{g}^{\gamma \delta} \partial_{\delta} \Omega}{\Omega}.
\end{equation}
So the Riemann tensor becomes
\begin{equation} 
\begin{aligned}
	\hat{R}^{\rho}_{\mu\sigma\nu}&={R}^{\rho}_{\mu\sigma\nu}-2{\nabla}_{[\mu}\hat{C}^{\rho}_{\sigma]\nu}+2\hat{C}^{\lambda}_{\nu]\mu}\hat{C}^{\rho}_{\sigma]\lambda}\\
	&={R}^{\rho}_{\mu\sigma\nu}+2\Omega^{-1}\delta_{[\mu}{\nabla}_{\sigma]}{\nabla}_{\nu}\Omega-2\Omega^{-1}\hat{g}^{\rho\lambda}\hat{g}_{\nu[\mu}{\nabla}_{\sigma]}{\nabla}_{\lambda}\Omega\\
	+&2\Omega^{-2}{\nabla}_{[\mu}\Omega\delta^{\lambda}_{\sigma]}{\nabla}_{\nu}\Omega-2\Omega^{-2}{\nabla}_{[\mu}\Omega\hat{g}_{\sigma]\nu}{\nabla}_{\xi}\Omega\\
	-&2\hat{g}_{\nu[\mu}\delta^{\rho}_{\sigma]}\hat{g}^{\lambda\xi}{\nabla}_{\xi}\Omega{\nabla}_{\lambda}\Omega
\end{aligned}
\end{equation}
and also for the Ricci tensor one has the following relation
\begin{equation} 
\begin{aligned}
	\label{eq:8}
	\hat{R}_{\mu\nu}=&{\nabla}_{\rho}\hat{C}^{\rho}_{\mu\nu}-{\nabla}_{\nu}\hat{C}^{\rho}_{\rho\mu}+\hat{C}^{\rho}_{\rho\lambda}\hat{C}^{\lambda}_{\mu\nu}-\hat{C}^{\rho}_{\mu\lambda}\hat{C}^{\lambda}_{\rho\nu}\\
	=&R_{\mu\nu}+(d-2)\Omega^{-2}{\nabla}_{\nu}\Omega{\nabla}_{\mu}\Omega\\
	-&(d-2)\Omega^{-2}\hat{g}_{\mu\nu}\hat{g}^{\rho\sigma}{\nabla}_{\rho}\Omega{\nabla}_{\sigma}\Omega\\
	-&(d-2)\Omega^{-1}{\nabla}_{\mu}{\nabla}_{\nu}\Omega-\Omega^{-1}\hat{g}_{\mu\nu}\hat{g}^{\rho\sigma}{\nabla}_{\rho}{\nabla}_{\sigma}\Omega.
\end{aligned}
\end{equation}
multiplying this relation by $\hat{g}_{\mu\nu}$ the Ricci scalar can be obtained 
\begin{equation}
	\hat{R}=R-2(d-1)\Omega^{-1}{\nabla}^{\nu}{\nabla}_{\nu}\Omega-(d-2)(d-1)\Omega^{-2}{\nabla}^{\nu}\Omega{\nabla}_{\nu}\Omega.
	\label{eq:9}
\end{equation}
Conformal transformation affects the null tetrad as \cite{ moreschi1986angular,moreschi1987general}
\begin{equation}
		\label{eq:5}
	\begin{aligned}
		&\hat{l}^{\mu}=\Omega^{-2} l^{\mu}, \\
		&\hat{m}^{\mu}=\Omega^{-1} m^{\mu}, \\
		&\hat{n}^{\mu}=n^{\mu}.
	\end{aligned}
\end{equation}
Note that the conformal factor is $\Omega=\rho$ in all these equations.
\section{\label{sec:level4} Einstein’s equations }
In this section we just try to give a short review on the work of Compère et al. \cite{compere2019lambda}.
Einstein’s equations $G_{\mu\nu}=\Lambda g_{\mu\nu}$, can be solved in the Bondi gauge. The aim of this section is to find $V/r$ and $\beta$ in de Sitter background. First  the expansion of $g_{AB}$ can be written as
\begin{equation}
	\label{eq:12}
	g_{A B}=r^{2} q_{A B}+r C_{A B}+D_{A B}+\frac{1}{r} E_{A B}+\frac{1}{r^{2}} F_{A B}+(C)\left(r^{-3}\right)
\end{equation}
where $q_{A B},\:C_{A B},\: D_{A B},\: E_{A B}$ and $F_{A B}$ are symmetric tensors which their components are arbitrary.  According to  \ref{eq:6} 
\begin{equation}
	g^{A B} \partial_{r} g_{A B}=4 / r.
\end{equation}
As in \cite{barnich2010aspects} $k_{AB}=\frac{1}{2} \partial_{r} g_{A B}, l_{A B}=\frac{1}{2} \partial_{u} g_{A B}$, and $n_{A}=\frac{1}{2} e^{-2 \beta} g_{A B} \partial_{r} U^{B}$ are defined, the indices of these parameters are uppering and lowering by $h_{AB}$.
Remember that as $R_{\mu\nu}=\Lambda g_{\mu\nu}$ then $R_{r r}=0$ that leads to \cite{compere2019lambda}
\begin{equation}
	\label{eq:10}
	\partial_{r} \beta=-\frac{1}{2 r}+\frac{r}{4} k_{B}^{A} k_{A}^{B}= \frac{1}{4 r^{3}} K_{B}^{A} K_{A}^{B}
\end{equation}
by expanding this relation one gets
\begin{equation}
	\begin{aligned}	
		\label{eq:13}	
\beta\left(u, r, x^{A}\right)=&\beta_{0}\left(u, x^{A}\right)+\frac{1}{r^{2}}\left[-\frac{1}{32} C^{A B} C_{A B}\right]\\
&+\frac{1}{r^{3}}\left[-\frac{1}{12} C^{A B} D_{A B}\right]\\
&+\frac{1}{r^{4}}[-\frac{3}{32} C^{A B} \mathcal{E}_{A B}\frac{1}{16} D^{A B} D_{A B}\\
&+\frac{1}{128}\left(C^{A B} C_{A B}\right)^{2}]+\mathcal{O}\left(r^{-5}\right)
	\end{aligned}
\end{equation}
which is in contrast to the result that is obtained for the flat case ($\frac{4}{r} \partial_{r} \beta=0$) where $\beta_{0}=0$.

On the other hand from $R_{r A}=0$, one can write
\begin{equation}
	\begin{aligned}
		\label{eq:11}
		G_{u r}+\Lambda g_{u r}=&R_{u r}+\frac{1}{2} g_{u r} R+\Lambda g_{u r}=\frac{1}{2} g_{u r} h^{A B} R_{A B}=0\\
		&\Rightarrow h^{A B} R_{A B}=0.
	\end{aligned}
\end{equation}
According to this, the radial dependence of $V/r$ can be fixed. Also from  $R_{r A}=0$, one can write
\begin{equation}
	\partial_{r}\left( \frac{r^{2}}{2} e^{-2 \beta} g_{A B} \partial_{r} U^{B}\right)=r^{2}\left(\partial_{r}-\frac{2}{r}\right) \partial_{A} \beta-D_{B} K_{A}^{B}
\end{equation}
where \ref{eq:10} is used. So
\begin{equation}
\begin{aligned}
		U^{A}=& U_{0}^{A}\left(u, x^{B}\right)+2 e^{2 \beta_{0}} \partial^{A} \beta_{0}\times \frac{1}{r}\\
		&-e^{2 \beta_{0}}\left[C^{A B} \partial_{B} \beta_{0}+\frac{1}{2} D_{B} C^{A B}\right] \times \frac{1}{r^{2}} \\
		&-\frac{2}{3} e^{2 \beta_{0}}[N^{A}-\frac{1}{2} C^{A B} D^{C} C_{B C}\\
		&+\left(\partial_{B} \beta_{0}-\frac{1}{3} D_{B}\right) D^{A B}-\frac{3}{16} C_{C D} C^{C D} \partial^{A} \beta_{0}] \times \frac{1}{r^{3}}\\
		&-\frac{2}{3} e^{2 \beta_{0}} D_{B} D^{A B} \frac{\ln r}{r^{3}}+o\left(r^{-3}\right),
		\end{aligned}
\end{equation}
where $D_A$ is the covariant derivative that is related to $q_{AB}$. Then from \ref{eq:11} 

\begin{equation}
	\label{eq19}
	\begin{aligned}
		\partial_{r} V=&-2 r\left(l+D_{A} U^{A}\right)+ e^{2 \beta} r^{2}[D_{A} D^{A} \beta\\
		&+\left(n^{A}-\partial^{A} \beta\right)\left(n_{A}-\partial_{A} \beta\right)-D_{A} n^{A}-\frac{1}{2} \mathcal{R}+\Lambda].
	\end{aligned}
\end{equation}
$\mathcal{R}$ is the Ricci scalar of the 2-metric $g_{AB}$ and has the following relation with the intrinsic curvature $K_{\mu\nu}$
\begin{equation}
\mathcal{R}=-\tilde{K}^2+\tilde{K}_{\mu\nu}\tilde{K}^{\mu\nu}-4\Omega^{-1}\tilde{K}.
\end{equation}
 The equation \ref{eq19} then leads to
\begin{equation}
	\label{eq16}
	\begin{aligned}
		\frac{V}{r}=& \frac{\Lambda}{3} e^{2 \beta_{0}} r^{2}-r\left(l+D_{A} U_{0}^{A}\right) \\
		&-e^{2 \beta_{0}}[\frac{1}{2}\left(\mathcal{R}+\frac{\Lambda}{8} C_{A B} C^{A B}\right)+2 D_{A} \partial^{A} \beta_{0}\\
		&+4 \partial_{A} \beta_{0} \partial^{A} \beta_{0}]-\frac{2 M}{r}+o\left(r^{-1}\right).
	\end{aligned}
\end{equation}
where \ref{eq:12}, \ref{eq:13} are used.   
 One can easily rewrite this equation using $\rho=1/r$ instead of $r$
\begin{equation}
	\label{eq17}
	\begin{aligned}
	V=& \frac{\Lambda}{3} e^{2 \beta_{0}} \rho^{-3}-\rho^{-2}\left(l+D_{A} U_{0}^{A}\right)\\
	&-\rho^{-1} e^{2\beta_{0}}[\frac{1}{2}\left(\mathcal{R}+\frac{\Lambda}{8} C_{A B} C^{A B}\right)+2 D_{A} \partial^{A} \beta_{0}\\
	&+4 \partial_{A} \beta_{0} \partial^{A} \beta_{0}]-2 M+o\left(\rho\right).
	\end{aligned}
\end{equation}
\section{\label{sec:level5}The de Sitter metric }
   The de Sitter line element in spherical coordinates has the form
   \begin{equation}
   	ds^2=-F(r)dt^2+F(r)dr^2+r^2d\Omega^2
   \end{equation}
where $F(r)=1-\Lambda r^2/3$ and $ d\Omega^2=d\theta^2+\sin \theta d\varphi^2$. One should consider that in this coordinate set, $t$ and $r$  become spacelike and timelike in $r>3/\Lambda$ zone.

 To find a  null tetrad  it is necessary to rewrite the line element
in advance null coordinates ( $u=t-r^*$, $r^*=\int dr/f(r)$ ). Then the line element  has the form \cite{saw2017behavior}
\begin{equation}
	ds^2=-F(r)du^2-2dudr+r^2q_{AB}dx^A dx^B
\end{equation}
where the unit 2-sphere is rewritten as $q_{AB}dx^A dx^B$. Also using the boundary gauge fixing $\beta=0,\: U^A=0, \:V/r=\Lambda r^2/3-1, h_{AB}=r^2 q_{AB}$ in Bondi-Sachs metric \ref{1}, one can obtain this result \cite{compere2019lambda}.
 \begin{equation}
	ds^2=\rho^{-2}(-F(\rho)\rho^2du^2+2dud\rho+q_{AB}dx^A dx^B)
\end{equation}
where $\rho=1/r$ is used instead of $r$. Then the matrix representation of the reversed metric is
\begin{equation}
	g^{\mu\nu}=
	\begin{bmatrix}
		0&\rho^{2}&0\\
		\rho^{2}&-F(\rho)&0\\
		0&0&q_{AB}
	\end{bmatrix}
\end{equation}
Accordingly, a null tetrad can be defined for this metric that satisfies the following relation \cite{newman1962approach} 	$g^{\mu\nu}=l^{\mu}n^{\nu}+l^{\nu}n^{\mu}-m^{\mu}\bar{m}^{\nu}-m^{\nu}\bar{m}^{\mu}$
\begin{align}
	\label{eq:4}
	& {l}^{\mu}=[0,-\rho^2,0,0],\\
	& {n}^{\mu}=[1,\frac{\rho^2 F(\rho)}{2},0,0]\notag,\\
	& {m}^{\mu}=[0,0,\frac{\rho q^A}{\sqrt{2}}].
\end{align}
 $q_A=2(1,i)/(1+\xi+\xi^*)$ is a complex dyad in stereographic coordinates ($\xi=e^{i\varphi}cot{\theta/2}$). $h_{AB}$ has the following relation with this dyad \cite{bishop2016gravitational}
 \begin{equation}
 	J:=h_{AB}q^{A}q^B/2
 \end{equation} 
 In the next section the form of this tetrad \ref{eq:4} in Bondi gauge is obtained.
 \section{\label{sec:level6}Coordinate transformation to the Bondi gauge}
General transformation can be used to write the null tetrad \ref{eq:4} in the Bondi gauge. The  transformation is written as follows \cite{bishop2016extraction}
\begin{equation}
\begin{aligned}
	&u \rightarrow \tilde{u}=u+\sum_{n=0} u_{n}\rho^n+\rho,\\
	&\rho \rightarrow \tilde{\rho}=\sum_{n=0} \omega_{n} \rho^{n+1},\\
	&x^{A} \rightarrow \tilde{x}^{A}=x^{A}+\sum_{n=0}x_{n}^{A}\rho^n.
	\end{aligned}
\end{equation}
$u_{n}, \omega_n, x_{n}^{A}$ are functions of $(u,x^A)$ and they can be obtained using
\begin{equation}
	\tilde{g}^{\alpha \beta}=\frac{\partial \tilde{x}^{\alpha}}{\partial x^{\mu}} \frac{\partial \tilde{x}^{\beta}}{\partial x^{\nu}} g^{\mu \nu},
\end{equation}
where $\tilde{g}^{\alpha \beta}$ is the metric in Bondi gauge and $g^{\mu\nu}$ is in Bondi-Sachs form. Thus $m^{\mu}$ in Bondi gauge has the form
\begin{equation}
\begin{aligned}
	\tilde{m}^{\alpha}=\frac{\partial \tilde{x}^{\alpha}}{\partial x^{\beta}} m^{\beta}=&((\partial_{B} u_{0}+\partial_{B} u_{1}\rho)\frac{\rho^2q^A}{\sqrt{2}},\partial_B\omega_0 \frac{\rho^3q^A}{\sqrt{2}}\\
	&,(1+\rho\partial_Bx_0^A)\frac{\rho^2q^A}{\sqrt{2}}).
\end{aligned}
\end{equation}
The other component of the tetrad is
\begin{equation}
\begin{aligned}
	\tilde{n}^{\alpha}=&(1-\frac{u_0}{2l^2}\rho+\partial_u u_0+O(\rho^2),-\frac{\omega_0}{2l^2}+\rho\partial_u\omega_0+O(\rho^2)\\
	&,-\frac{x_0^A}{l^2}\rho\partial_u x_0^A).
	\end{aligned}
	\end{equation}
Also the metric in Bondi gauge has the relation $\tilde{g}_{\alpha \beta}=\tilde{\rho}^{-2} \hat{\tilde{g}}_{\alpha \beta}$ and one can obtain the relation between $\hat{g}_{\alpha \beta}$ and $\hat{\tilde{g}}_{\alpha \beta}$ \cite{bishop2016extraction}
\begin{equation}
	\begin{aligned}
	\hat{g}_{\alpha \beta}=&\rho^{2} g_{\alpha \beta}\\
	=&\rho^{2} \frac{\partial \tilde{x}^{\gamma}}{\partial x^{\alpha}} \frac{\partial \tilde{x}^{\delta}}{\partial x^{\beta}} \tilde{g}_{\gamma \delta}\\
	=&\frac{\rho^{2}}{\tilde{\rho}^{2}} \frac{\partial \tilde{x}^{\gamma}}{\partial x^{\alpha}} \frac{\partial \tilde{x}^{\delta}}{\partial x^{\beta}} \hat{\tilde{g}}_{\gamma \delta}\\
	=&\frac{1}{\omega_0^{2}} \frac{\partial \tilde{x}^{\gamma}}{\partial x^{\alpha}} \frac{\partial \tilde{x}^{\delta}}{\partial x^{\beta}} \hat{\tilde{g}}_{\gamma \delta}+\mathcal{O}(\rho).
		\end{aligned}
\end{equation} 
Similarly to 	\ref{eq:4}
\begin{equation}
	\begin{aligned}
		&\hat{\tilde{l}}^{\mu}=\Omega^{-2} \tilde{l}^{\mu}, \\
		&\hat{\tilde{m}}^{\mu}=\Omega^{-1} \tilde{m}^{\mu}, \\
		&\hat{\tilde{n}}^{\mu}=\tilde{n}^{\mu}.
	\end{aligned}
\end{equation}

\section{\label{sec:level7} The news tensor }
It is better to use the coordinate independent notion of the news tensor that is defined using the null tetrad \cite{penrose1963asymptotic, bishop1997high}
	\begin{equation}
		\label{eq:23}
		N_{\mu\nu}=\lim _{\tilde{\rho} \rightarrow 0} \frac{\hat{\tilde{m}}^{\alpha} \hat{\tilde{m}}^{\beta} \hat{\tilde{\nabla}}_{\alpha} \hat{\tilde{\nabla}}_{\beta} \tilde{\rho}}{\tilde{\rho}}
	\end{equation}
where
\begin{equation}
	\hat{\tilde{\nabla}}_{\alpha} \hat{\tilde{\nabla}}_{\beta} \tilde{\rho}=\tilde{\nabla}_{\alpha} \tilde{\nabla}_{\beta} \tilde{\rho}+\frac{\hat{\tilde{g}}_{\alpha b} \hat{\tilde{g}}^{11}-2 \delta_{\alpha}^{1} \delta_{\beta}^{1}}{\tilde{\rho}}
\end{equation}
So taking $\tilde{\rho}\simeq\omega_0\rho$, the news tensor \ref{eq:23} becomes \begin{equation}
	\tilde{m}^{\alpha} \tilde{m}^{\beta} \tilde{\nabla}_{\alpha} \tilde{\nabla}_{\beta} \tilde{\rho}-m^{\alpha} m^{\beta} \nabla_{\alpha} \nabla_{\beta} \tilde{\rho}=0
\end{equation}

\begin{equation}
	\begin{aligned}
		N_{\mu\nu}=&\lim_{\tilde{\rho}\rightarrow 0}
\frac{	\overbrace{	\tilde{m}^{\alpha} \tilde{m}^{\beta}\tilde{\nabla}_{\alpha} \tilde{\nabla}_{\beta} \tilde{\rho}}^{{m}^{\alpha}{m}^{\beta}{\nabla}_{\alpha} {\nabla}_{\beta} (\omega_0 \rho)}+\tilde{m}^{\alpha} \tilde{m}^{\beta}\frac{\hat{\tilde{g}}_{\alpha b} \hat{\tilde{g}}^{11}-2 \delta_{\alpha}^{1} \delta_{\beta}^{1}}{\tilde{\rho}}}{\tilde{\rho}}\\
		=&\lim_{\rho\rightarrow 0}
	\frac{1}{\omega_0\rho}\left[m^{\alpha} m^{\beta}({\nabla}_{\alpha} {\nabla}_{\beta}(\rho \omega_0)\right.\\
		&\left.-\hat{g}_{\alpha \beta}\left(\frac{\hat{g}^{11} \omega}{\rho}+\hat{g}^{1 \gamma} \partial_{\gamma} \omega\right)-\frac{2 \rho \partial_{\alpha} \omega \partial_{\beta} \omega}{\omega})\right]
	\end{aligned}
\end{equation}
The result is 
\begin{equation}
	\begin{aligned}
		N_{\mu\nu}&=\lim _{\tilde{\rho} \rightarrow 0} \frac{\hat{\tilde{m}}^{\alpha} \hat{\tilde{m}}^{\beta} \hat{\nabla}_{\alpha} \hat{\nabla}_{\beta} \tilde{\rho}}{\tilde{\rho}}\\
		&=\frac{1}{\omega^{2}}\left[-\lim _{\rho \rightarrow 0} \frac{q^{A} q^{B} \hat{C}_{A B}^{1}}{\rho}\right.\\
		&\left.+q^{A} q^{B}(\frac{\partial_{A} \partial_{B} \omega}{\omega}-\frac{\hat{C}_{A B}^{\gamma} \partial_{\gamma} \omega}{\omega}\right.\\
		&\left.-\frac{\partial_{\rho} \hat{g}_{A B} e^{-2 \beta} (V\rho^2+\rho)}{2}-\frac{2 \partial_{A} \omega_0 \partial_{B} \omega_0}{\omega_0^{2}})\right]
	\end{aligned}
\end{equation}
Using \ref{eq17}, one has 

\begin{equation}
	\begin{aligned}
		N_{\mu\nu}&=\lim _{\tilde{\rho} \rightarrow 0} \frac{\hat{\tilde{m}}^{\alpha} \hat{\tilde{m}}^{\beta} \hat{\nabla}_{\alpha} \hat{\nabla}_{\beta} \tilde{\rho}}{\tilde{\rho}}\\
		&=\frac{1}{\omega^{2}}\left[-\lim _{\rho \rightarrow 0} \frac{q^{A} q^{B} \hat{C}_{A B}^{1}}{\rho}\right.\\
		&\left.+q^{A} q^{B}(\frac{\partial_{A} \partial_{B} \omega}{\omega}-\frac{\hat{C}_{A B}^{\gamma} \partial_{\gamma} \omega}{\omega}\right.\\
		&\left.-  \frac{\partial_{\rho}\hat{g}_{A B} e^{-2 \beta}}{2}\left[\frac{\Lambda}{3} e^{2 \beta_{0}} \rho^{-1}-\left(l+D_{A} U_{0}^{A}\right)\right]\right.\\
		&\left. -\rho e^{2 \beta_{0}}[\frac{1}{2}\left(\mathcal{R}+\frac{\Lambda}{8} C_{A B} C^{A B}\right)+2 D_{A} \partial^{A} \beta_{0}\right.\\
		&\left.+4 \partial_{A} \beta_{0} \partial^{A} \beta_{0}]\right.\\
		&\left.+\rho-2 M \rho^{2}+o\left(\rho^{3}\right)) \}-\frac{2 \partial_{A} \omega_0 \partial_{B} \omega_0}{\omega_0^{2}})\right].
	\end{aligned}
\end{equation}
This is the Bondi news tensor in de Sitter background, which is different from the Bondi news in $\Lambda=0$ case where $\beta_0$ was set to zero. 
\section{Conclusion}
In this work a new result for Bondi news in de Sitter space-time is obtained. This result is acquired  by using a null tetrad for the de Sitter metric in Bondi gauge. Changing the Minkowski background to the de Sitter background insert new parts to the known formula.

The approach used in this article is to write the de Sitter metric in Bondi-Sachs form that reduces to the ordinary form of the de Sitter metric by choosing $\beta=0, U^{A}=0, V / r=\left(\Lambda r^{2} / 3\right)-1, g_{A B}= r^2d\Omega^2$. Ergo, the asymptotic expansion of  $ g_{A B}$ in power of $r$ is used to solve Einstein's equations and obtain Bondi-Sachs variables $V$ and $\beta$.

 Then the de Sitter line element is written  in "coordinate-like Eddington-Finkelstein-type" coordinates \cite{saw2017mass} and $\rho=1/r$ is used in order to find a null tetrad. Finally the tetrad is obtained in $(u,\rho,x^A)$ coordinates, which reduces exactly to asymptotically flat case when $\Lambda$ goes to zero. 

Considering the form of this tetrad affected by the general transformation to the Bondi gauge we calculate the Bondi news. To obtain this result, a notion of Bondi news is used, based on the null tetrad.
 One can see that how a non-zero cosmological constant can add new terms to Bondi news tensor obtained for the flat case.  These new terms
 are so small that they are neglectable.

\nocite{*}

\bibliography{apssamp}
\end{document}